\documentclass[trackchanges,twocolumn]{aastex701}

\usepackage[utf8]{inputenc}
\usepackage{amsmath,amsxtra,amssymb,latexsym,amscd,amsthm}
\usepackage{mathrsfs}
\usepackage[mathscr]{eucal}
\usepackage{mathrsfs}
\usepackage{makeidx}
\usepackage{soul}
\usepackage{graphicx}
\usepackage{subfigure}
\usepackage{booktabs}
\usepackage{multirow}
\usepackage{threeparttable}
\usepackage{hyperref}
\usepackage{makecell}

\shorttitle{Dust Concentration in Turbulent and Windy Protoplanetary Disks}
\shortauthors{Huang \& Bai}
\graphicspath{{./}{figure/}}

\begin{document}

\title{Planetesimal Formation under Realistic Gas Dynamics: Dust Concentration in Turbulent and Windy Protoplanetary Disks}

\author[orcid=0000-0002-7575-3176, gname='Pinghui', sname='Huang']{Pinghui Huang}
\affiliation{CAS Key Laboratory of Planetary Sciences, Purple Mountain Observatory, Chinese Academy of Sciences, Nanjing 210023, People’s Republic of China}
\email[show]{phhuang@pmo.ac.cn}

\author[orcid=0000-0001-6906-9549, gname='Xue-Ning', sname='Bai']{Xue-Ning Bai}
\affiliation{Institute for Advanced Study, Tsinghua University, Beijing 100084, People’s Republic of China}
\affiliation{Department of Astronomy, Tsinghua University, Beijing 100084, People’s Republic of China}
\email[show]{xbai@mail.tsinghua.edu.cn}

\begin{abstract}
Planetesimals are a key intermediate stage of planet formation and originate from dust concentration in protoplanetary disks, which is strongly regulated by gas dynamics. In this study, we use the multifluid dust module of Athena++ to investigate dust concentration in 2D axisymmetric global non-ideal MHD (ambipolar diffusion, AD) simulations of outer disks. Our simulations simultaneously launch large-scale MHD disk winds and resolve both streaming instability (SI) and vertical shear instability (VSI). We systematically explore the effects of magnetic field strength, cooling timescale, ambipolar Els\"asser number, and dust size on dust concentration. We find that strong dust clumping persists across a broad range of magnetized and wind-launching disk conditions. MHD winds also enhance dust clumping by modifying background conditions in the following ways: (1) secular gas depletion by MHD wind mass loss, and (2) dust trapping in zonal flows as a result of magnetic flux concentration (which occurs in more strongly magnetized cases). Overall, our results demonstrate that strong dust clumping can persist in global magnetized disks with AD and large-scale MHD winds, leading to conditions favorable for planetesimal formation.
\end{abstract}

\section{Introduction}~\label{sec:introductions}

Planetesimal formation is a critical yet still not fully understood stage in the bottom-up scenario of planet formation in protoplanetary disks (PPDs). In this picture, micron-sized dust grains grow through coagulation into pebble-sized particles, which must further concentrate to form kilometer-sized planetesimals. However, this growth is strongly hindered by the ``meter-size'' barrier, as solids of this size experience rapid radial drift toward the central star~\citep{Birnstiel2024}. Streaming instability~\cite[SI;][]{YoudinJohansen2007,JohansenYoudin2007} has been proposed as a promising mechanism to overcome this barrier by aerodynamically concentrating pebbles into dense clumps, which may gravitationally collapse into planetesimals once the local dust density exceeds the Roche density. Dust can also accumulate in pressure bumps, such as at the edges of planet-opened gaps~\citep{ZhuStone2014}, within zonal flows generated by disk instabilities~\citep{Johansen2009}, or inside long-lived anticyclonic vortices~\citep{BargeSommeria1995}, providing alternative pathways for planetesimal formation. However, most previous numerical studies of dust concentration adopt simplified gas dynamics, including laminar disks without external turbulence~\citep{LiYoudin2021,LiChiang2025}, $\alpha$-viscosity prescription~\citep{ChenLin2020,Umurhan2020}, or artificially forced turbulence~\citep{GoleSimon2020,LimSimon2024}. Whether dust concentration behaves similarly under more realistic disk dynamics remains an open question.

The gas dynamics of PPDs is closely linked to disk evolution and plays a fundamental role in dust transport and concentration. Because of the low ionization levels in PPDs, the classical magnetorotational instability~\cite[MRI;][]{BalbusHawley1998} is often suppressed or damped by non-ideal magnetohydrodynamic (MHD) effects~\citep{Bai2011MRI}, including Ohmic resistivity, Hall effect, and ambipolar diffusion (AD). These non-ideal MHD effects significantly modify the MRI and lead to more complex disk dynamics. Instead, several hydrodynamical instabilities, including vertical shear instability~\cite[VSI;][]{Nelson2013,BarkerLatter2015}, convective overstability~\cite[COS;][]{Lyra2014,Latter2016}, and zombie vortex instability~\cite[ZVI;][]{Marcus2015,Marcus2016}, may operate in different disk regions and drive turbulence and disk evolution~\citep{FromangLesur2019,LyraUmurhan2019,Lesur2023,Bai2026}. In addition, magnetized disk winds have been suggested to play a major role in driving disk accretion and shaping disk structures, potentially producing accretion rates consistent with observations~\citep{BaiStone2013Wind}.

Recent studies have begun to explore dust dynamics and planetesimal formation in more realistic disk environments. For example, dust clumping driven by SI has been investigated in MRI turbulence with AD~\citep{XuBai2022,Eriksson2026}. The interaction between VSI turbulence and SI has also been explored using 2D axisymmetric simulations~\citep{SchaferJohansen2020,SchaferJohansen2022,HuangBai2025I}, while global 3D simulations have shown that the interplay between multiple instabilities can lead to strong dust concentration~\citep{HuangBai2025II}. Meanwhile, magnetized disk winds may modify the radial transport and accumulation of solids, and magnetic flux concentration (MFC) generated by AD can further enhance dust concentration through zonal flows~\citep{RiolsLesur2020,HuLiZhu2022,HsuLi2025}. These studies suggest that the efficiency and locations of planetesimal formation may strongly depend on the underlying gas dynamics.

To this end, we investigate dust concentration in a self-consistent turbulent and wind-launching environment based on first-principles modeling. Using mesh refinement in Athena++, we simultaneously capture large-scale wind motions, intermediate-scale VSI turbulence/MHD activity and zonal flows induced by MFC, and small-scale SI dust clumping. This approach allows us to explore how dust concentrates under realistic disk dynamics and to assess its implications for planetesimal formation in PPDs. The outline of this paper is as follows. Section~\ref{sec:setups} describes the numerical setups. Section~\ref{sec:results} presents the numerical results. Section~\ref{sec:summary} provides conclusions and discussions.

\section{Numerical Setups}~\label{sec:setups}

\subsection{Equations and Mesh}~\label{subsec:equations}

We employ the magnetohydrodynamics code Athena++~\citep{Stone2020,AthenaPP2024}, together with its multifluid dust module\footnote{https://github.com/PinghuiHuang/athena-multifluid-dust}~\citep{HuangBai2022}, to investigate dust–gas interactions in the presence of non-ideal MHD effects. Our numerical setup is similar with that of~\cite{CuiBai2020}, but includes dust species. Simulations in this study are set to be scale free. The governing equations for the dust, gas, and magnetic field used in our simulations are summarized below.

\begin{equation}
  \frac{\partial \rho_{\text{d}}}{\partial t} +\nabla \cdot \left(\rho_{\text{d}} \mathbf{v}_{\text{d}}\right) = 0
\label{eq:dust_con}
\end{equation}

\begin{equation}
\begin{aligned}
  \frac{\partial (\rho_{\text{d}} \mathbf{v}_{\text{d}})}{\partial t} + \nabla\cdot\left(\rho_{\text{d}}\mathbf{v}_{\text{d}} \mathbf{v}_{\text{d}} \right) = \rho_{\text{d}} \nabla \Phi + \rho_\text{d} \Omega_\text{K} \frac{(\mathbf{v}_\text{g} - \mathbf{v}_\text{d})}{St}
\end{aligned}
\label{eq:dust_mom}
\end{equation}

\begin{equation}
\frac{\partial \rho_{\text{g}}}{\partial t} + \nabla \cdot \left(\rho_{\text{g}} \mathbf{v}_{\text{g}}\right) =0
\label{eq:gas_con}
\end{equation}

\begin{equation}
\begin{aligned}
  \frac{\partial (\rho_{\text{g}}\mathbf{v}_{\text{g}})}{\partial t} + & \nabla\cdot\left(\rho_{\text{g}}\mathbf{v}_{\text{g}} \mathbf{v}_{\text{g}} - \frac{\mathbf{B}\mathbf{B}}{4\pi} + \frac{|\mathbf{B}|^2}{8\pi}\mathsf{I} + P\mathsf{I}\right)
  \\ &= \rho_{\text{g}} \nabla \Phi +\rho_\text{d} \Omega_\text{K} \frac{(\mathbf{v}_\text{d} - \mathbf{v}_\text{g})}{St}
\end{aligned}
\label{eq:gas_mom}
\end{equation}

\begin{equation}
\begin{aligned}
  \frac{\partial E}{\partial t} &+ \nabla \cdot \left[ \left(E + \frac{|\mathbf{B}|^2}{8\pi} + P\right)\mathbf{v}_\text{g} - \frac{\mathbf{B(\mathbf{B}\cdot \mathbf{v}_\text{g})}}{4\pi} + \mathbf{S_\text{AD}}\right]\\
&=\rho_\text{g}\mathbf{v}_\text{g} \cdot \nabla \Phi + \rho_\text{d}\Omega_\text{K} \mathbf{v}_\text{g} \cdot \frac{\left(\mathbf{v}_\text{d} - \mathbf{v}_\text{g}\right)}{St}\\
&+ \rho_\text{d}\Omega_\text{K}\frac{\left(\mathbf{v}_{\text{d}} - \mathbf{v}_{\text{g}}\right)^2}{St} - \frac{\rho_\text{g}\Omega_\text{K}}{\gamma - 1} \frac{\left(T - T_\text{init}\right)}{\beta_\text{th}}
\end{aligned}
\label{eq:gas_erg}
\end{equation}

\begin{equation}
\begin{aligned}
  \frac{\partial \mathbf{B}}{\partial t} = \nabla\times \left(\mathbf{v}_\text{g} \times \mathbf{B} - c \mathbf{E_\text{AD}}\right)
\end{aligned}
\label{eq:B_induction}
\end{equation}

Equations~\ref{eq:dust_con} and~\ref{eq:dust_mom} are the continuity and momentum equations for dust, while Equations~\ref{eq:gas_con}--\ref{eq:gas_erg} describe the continuity, momentum, and energy evolution for gas. Equation~\ref{eq:B_induction} is the induction equation governing magnetic field evolution. Here, $\rho$, $\mathbf{v}$, $P$, and $E$ denote density, velocity, thermal pressure, and total gas energy, respectively, and $\mathbf{B}$ denotes magnetic field. The subscripts ``$\mathrm{d}$'' and ``$\mathrm{g}$'' refer to dust and gas quantities. In this study, we include only one dust species to facilitate a systematic investigation of the dust--gas interaction. We adopt an adiabatic equation of state with $\gamma = 1.4$. The gravitational potential is given by $\Phi = GM/r$, and we adopt $G=M=1$ in this work. To regulate the thermal structure of the disk, we include a thermal relaxation term in Equation~\ref{eq:gas_erg}, which relaxes the gas temperature toward its initial value $T_\mathrm{init}$ (see Section~\ref{subsec:ICandBC}) over a cooling timescale $\tau_\mathrm{cool}$. The dimensionless thermal relaxation timescale is defined as $\beta_\mathrm{th} \equiv \tau_\mathrm{cool}\Omega_\mathrm{K}$ and $\Omega_\mathrm{K}\equiv\sqrt{GM/R^3}$ is the Keplerian frequency. In reality, small dust grains in PPDs are important cooling agents, an effect that is not explicitly modeled here because we adopt a simple $\beta$-cooling prescription. Our focus is instead on larger grains that are marginally coupled to the gas and participate in planetesimal formation near the midplane. Since our goal is to capture the effects of self-consistent gas dynamics on dust evolution, we do not include any explicit prescriptions for turbulent viscosity or dust diffusion.

In the outer regions of PPDs ($\gtrsim10\,\mathrm{AU}$), non-ideal MHD effects are expected to be dominated by AD~\citep{Bai2011MRI}, where electrons and ions are coupled to magnetic fields, which can drift relative to the neutrals through ion-neutral friction. We therefore include AD as the only non-ideal effect in this work, leaving Ohmic resistivity and Hall effect~\citep{Bai2014HallI,Bai2015HallII} for future studies. The imperfect coupling of magnetic field to the gas (neutrals) introduces the AD electric field
\begin{equation}
  \mathbf{E}_\mathrm{AD} = \frac{4\pi}{c^2} \eta_\mathrm{A} \mathbf{J}_{\perp}\ .
\end{equation}
Here, the current density is defined as $\mathbf{J}=\nabla\times\mathbf{B}$ and $\mathbf{J}_{\perp}$ denotes the component of the current density perpendicular to the magnetic field. The corresponding Poynting flux, $\mathbf{S}_\mathrm{AD}=c(\mathbf{E}_\mathrm{AD}\times\mathbf{B})/4\pi$, represents the transport of magnetic energy associated with this ion-neutral drift. The AD diffusivity denotes $\eta_\mathrm{A}$, which generally scales as $\eta_\mathrm{A}\propto |\mathbf{B}|^2/\rho_\mathrm{g}^2$. The importance of magnetic coupling relative to AD is characterized by the Els\"asser number,
\begin{equation}
Am \equiv \frac{v_A^2}{\eta_\mathrm{A}\Omega_\mathrm{K}}
\end{equation}
where $v_A \equiv |\mathbf{B}|/\sqrt{4\pi\rho_\mathrm{g}}$ is the Alfv\'en speed. Similarly, both $v_A^2$ and $\eta_\mathrm{A}$ scale as $|\mathbf{B}|^2$, the Els\"asser number, $Am$, does not explicitly depend on the magnetic field strength.

We consider one dust species in the present study. The dust size is characterized by the Stokes number, defined as $\mathrm{St}\equiv \tau_\mathrm{st}\Omega_\mathrm{K}$, where $\tau_\mathrm{st}$ is the dust--gas drag stopping time. Dust and gas are coupled through aerodynamic drag terms appearing in both the momentum and energy equations. In Equation~\ref{eq:gas_erg}, the source terms associated with dust–gas interaction represent the work done by drag forces and the frictional heating generated by dust–gas relative motion.  In this study, we assume that all dissipated drag energy is converted into the internal energy of the gas. The frictional heating caused by mutual dust--gas drag is negligible compared with compressional heating~\citep{Segretain2024}. Moreover, most of our simulations adopt instantaneous cooling, which rapidly removes any thermal effects. We therefore expect dust--gas frictional heating to have little effect on the main conclusions of this study. We neglect self-gravity in this work and leave its effects for future investigation.

We perform our simulations in spherical--polar coordinates $(r,\theta,\phi)$ using Athena++, adopting a 2D axisymmetric setup. For clarity, we present the results in cylindrical coordinates $(R,\phi,z)$ throughout this paper. The coordinate transformation is given by $R = r \sin\theta$ and $z = r \cos\theta$. We adopt the HLLD Riemann solver for the gas and a custom solver for the dust~\citep{HuangBai2022}. Time integration is performed using the VL2 integrator, with a super-time-stepping scheme to accelerate the treatment of AD.

The computational domain in our 2D axisymmetric simulations spans $r\in[1,100]$ and $\theta\in[0,\pi]$, resolved with $1536\times1024$ grid cells at the root-grid level. The radial grid is logarithmically spaced with a constant ratio of $\Delta r_{i+1}/\Delta r_i=1.003$, while the $\theta$ grid is uniformly spaced, yielding $\Delta r:r\Delta\theta \approx 1$ throughout the radial domain. In this study, $H_\mathrm{g}\propto R$ (see Section~\ref{subsec:ICandBC}), while both $\Delta r$ and $r\Delta\theta$ also scale approximately linearly with radius. Here, $H_\mathrm{g}\equiv c_s/\Omega_\mathrm{K}$ is the gas scale height, with $c_s\equiv\sqrt{P/\rho_\mathrm{g}}$ denoting the sound speed. Consequently, the root-grid resolution remains nearly constant at approximately $33$ cells per $H_\mathrm{g}$ in both the radial and meridional directions throughout the disk region.

To better resolve the dust--gas dynamics near the disk midplane, we employ three levels of mesh refinement centered on the midplane. Refinement is activated at $t=1000\,P_0$, with the finest grid covering the region $r\in[3,10]$ and $\theta\in[\pi/2-0.025,\ \pi/2+0.025]$. With three levels of refinement, each doubling the resolution, the effective resolution within the refined region reaches approximately $267$ cells per $H_\mathrm{g}$ in both the radial and meridional directions. Based on the resolution studies presented in~\cite{YangJohansen2014} and~\cite{LiYoudin2018}, a resolution of approximately $160$--$320$ cells per $H_\mathrm{g}$ is expected to resolve the relevant growing SI modes in stratified simulations.

\subsection{Initial and Boundary Conditions}~\label{subsec:ICandBC}

\begin{table*}
\centering
\scriptsize
\setlength{\tabcolsep}{0.1pt}
\renewcommand{\arraystretch}{0.6}
\caption{Problem setup for 2D axisymmetric runs}
\begin{tabular}{cccccc}
\toprule
\hline
Model &
\makecell[l]{Initial strength of poloidal \\ field in midplane \\ $\beta_{\rm pl,mid,init}$} &
\makecell[l]{Els\"asser number of \\ AD in disk region \\ $Am_{\rm disk}$} &
\makecell[l]{Thermal cooling \\ in disk region \\ $\beta_{\rm th,disk}$} &
\makecell[l]{Dust Stokes \\ number \\ $St$} &
\makecell[l]{Initial dust--gas \\ surface density \\ ratio $Z_\mathrm{d}$} \\
\midrule
Fid & $10^{4}$ & 0.3 & $ 10^{-6}$ &   0.1  &   0.01 \\
B3  & $10^{3}$ & 0.3 & $ 10^{-6}$ &   0.1  &   0.01 \\
B5  & $10^{5}$ & 0.3 & $ 10^{-6}$ &   0.1  &   0.01 \\
Am3 & $10^{4}$ & 3   & $ 10^{-6}$ &   0.1  &   0.01 \\
T1  & $10^{4}$ & 0.3 &  $1$    &   0.1  &   0.01 \\
S2  & $10^{4}$ & 0.3 & $ 10^{-6}$ &   0.01 &   0.02 \\
\bottomrule
\label{table:runs}
\end{tabular}
\end{table*}

The initial gas density $\rho_\mathrm{g,init}$ and temperature $T_\mathrm{init}$ profiles are prescribed as
\begin{equation}
\label{eq:rho_init}
\rho_\mathrm{g,init} = \rho_0 \left( \frac{r}{R_0} \right)^{-q_\text{D}} f(\theta)
\end{equation}
and
\begin{equation}
\label{eq:T_init}
T_\mathrm{init} = \frac{P_\mathrm{init}}{\rho_\mathrm{g,init}} = \frac{GM}{r} h^2(\theta)
\end{equation}
where $\rho_0=1$, $R_0=1$ and $P_\mathrm{init}$ is the initial thermal pressure. The radial density profile follows a power law with index $q_\text{D}=2.25$, corresponding to an initial surface density profile of $\Sigma_{\mathrm{g,init}}\propto R^{-1}$. The function $h(\theta)$ specifies the prescribed aspect ratio, which is radially independent for a self-similar equilibrium disk solution in which the radial and meridional structures can be solved separately (i.e., $T_\mathrm{init}\propto R^{-1}$). This choice makes the aspect ratio within the disk region constant (i.e., $H_\mathrm{\mathrm{g}} \propto R$) and maintains an approximately constant number of grid cells per $H_\mathrm{g}$ with radius. In particular, the function $f(\theta)$ is determined implicitly from \citep{BaiStone2017Hall}:
\begin{equation}
\label{eq:f_theta}
\frac{d\ln[f(\theta)h^2(\theta)]}
{d\ln(\sin\theta)} = \frac{1}{h^2(\theta)} (q_\mathrm{D}+1)
\end{equation}
The initial gas velocities are initialized as $v_{\mathrm{g},r,\mathrm{init}}=0$, $v_{\mathrm{g},\theta,\mathrm{init}}=0$ and while the initial azimuthal velocity is determined by radial force balance:
\begin{equation}
\label{eq:vphi_init}
v_{\mathrm{g},\phi,\mathrm{init}}^2 = \frac{GM}{r} \left[ 1-(q_\mathrm{D}+1)h^2(\theta) \right]
\end{equation}

We prescribe the aspect-ratio function $h(\theta)$ as
\begin{equation}
\label{eq:h_theta}
h(\theta) = h_\mathrm{disk} + \frac{1}{2} \left( h_\mathrm{wind}-h_\mathrm{disk} \right) \left[ \tanh \frac{2\left(\delta\theta - \theta_\mathrm{tr}\right)}{h_\mathrm{disk}} +1 \right]
\end{equation}
where we adopt $h_{\mathrm{disk}}=H_\mathrm{g}/R=0.1$ in the disk region and $h_{\mathrm{wind}}=0.5$ in the wind region. Here, $\delta\theta \equiv |\theta-\pi/2|$ denotes the absolute latitude. The transition between $h_\mathrm{disk}$ and $h_\mathrm{wind}$ occurs at $z_\mathrm{tr} = r\theta_\mathrm{tr} = 3.5H_\mathrm{g}$ above and below the midplane~\citep{CuiBai2021}, where $\theta_\mathrm{tr}=0.35$ is the angular offset from the midplane ($\theta=\pi/2$, $z=0$). This transition is introduced to mimic the temperature increase near the disk surface due to stellar far-UV (FUV)/X-ray (XUV) irradiation, as described by Equations~\ref{eq:f_theta}--\ref{eq:h_theta}. The adopted value of $h_\mathrm{disk}$ is approximately comparable with the typical aspect ratio of outer disks, while $h_\mathrm{wind}$ is chosen for numerical convenience~\citep{CuiBai2021,CuiBai2022}.

Physically, the same XUV irradiation also substantially boosts the ionization fraction, driving the gas toward the ideal-MHD regime~\citep{Perez-BeckerChiang2011}. This ionization transition naturally facilitates disk-wind launching. Accordingly, we prescribe a vertical profile of the ambipolar Els\"asser number $Am$ that smoothly transitions from a non-ideal MHD regime in the disk to a nearly ideal-MHD regime in the wind region in a similar fashion:
\begin{equation}
\begin{aligned}
Am &= Am_\mathrm{disk} \\
\quad +&\frac{1}{2} \left(Am_\mathrm{wind}-Am_\mathrm{disk}\right) \left[ \tanh \frac{ 2(\delta\theta-\theta_\mathrm{tr}) } {h_\mathrm{disk}}  +1 \right]
\end{aligned}
\end{equation}
where $Am_\mathrm{disk}$ and $Am_\mathrm{wind}$ are the Els\"asser numbers in the disk and wind regions, respectively. In this work, we fix $Am_\mathrm{wind}=100$, corresponding to a nearly ideal-MHD regime in the wind region. The disk Els\"asser number $Am_\mathrm{disk}$ is set to either $0.3$ or $3$, representing cases with stronger and weaker AD~\citep{Bai2011Grain}, respectively.

The poloidal magnetic field is initialized through an azimuthal vector potential
$A_\phi(r,\theta)$~\citep{Zanni2007,BaiStone2017Hall}:
\begin{equation}
  A_\phi(r,\theta) = \frac{2B_{z0}R_0} {3-q_\mathrm{D}} \left( \frac{R}{R_0} \right)^{\frac{1-q_\mathrm{D}}{2}} \left[ 1+ \left( \frac12\tan\theta \right)^{-2} \right]^{-\frac{5}{8}}
\end{equation}
This vector potential is chosen such that the resulting magnetic field has a vertical component in the disk midplane given by
\begin{equation}
\mathbf{B}_{\mathrm{mid}} = B_{z0} \left( \frac{R}{R_0} \right)^{-(q_\mathrm{D}+1)} \hat{z}
\end{equation}
where $B_{z0}$ is the vertical magnetic field strength at the reference radius $R_0\equiv1$. This radial profile is selected to maintain a constant initial midplane plasma beta parameter:
\begin{equation}
\beta_{\mathrm{pl,mid,init}} = \frac{8\pi P_\mathrm{init}} {|\mathbf{B}_{\mathrm{mid}}|^2}
\end{equation}

The thermal relaxation parameter $\beta_\mathrm{th}$ in Equation~\ref{eq:gas_erg} is prescribed to take different values in the disk and wind regions, denoted by $\beta_\mathrm{th,disk}$ and $\beta_\mathrm{th,wind}$, respectively, with a smooth transition between the two described by a hyperbolic tangent function. We fix $\beta_\mathrm{th,wind}=10^{-6}$, corresponding to effectively instantaneous thermal relaxation in the wind region. The disk thermal relaxation parameter is set to either $\beta_\mathrm{th,disk}=10^{-6}$ or $1$, allowing us to conveniently control the disk thermal response.

The initial dust density is set as $\rho_{\mathrm{d,init}}=Z_\mathrm{d}\rho_{\mathrm{g,init}}$, where $Z_\mathrm{d}$ denotes the initial dust--gas surface density ratio (metallicity). In this study, we adopt $Z_\mathrm{d}=0.01$ or $0.02$. As the initial dust--gas density ratio is vertically uniform, the dust density follows the same vertical distribution as the gas, corresponding to an initial dust scale height of $H_\mathrm{d,init}=H_\mathrm{g}$. Since the initial dust--gas ratio is small, dust feedback has little influence on the gas dynamics at early times. We therefore initialize the dust velocities to match the gas velocities, i.e., $v_{\mathrm{d},r,\mathrm{init}} = v_{\mathrm{g},r,\mathrm{init}}$, $v_{\mathrm{d},\theta,\mathrm{init}} = v_{\mathrm{g},\theta,\mathrm{init}}$, and $v_{\mathrm{d},\phi,\mathrm{init}} = v_{\mathrm{g},\phi,\mathrm{init}}$.

Due to the rapid decrease in gas density near the polar regions during the simulations, we implement dust-density damping zones near the north and south poles to suppress numerical artifacts caused by excessively low gas density resulting from severe gas depletion. Within these regions ($\theta<0.4$ or $\theta>\pi-0.4$), the dust density is smoothly relaxed toward a floor value of $10^{-4}Z_\mathrm{d}\rho_\mathrm{g}$ throughout the simulations. The total dust mass removed by the polar dust damping zones remains below $0.1\%$ of the total dust mass throughout the simulations.

Similar to~\cite{CuiBai2020}, we adopt a polar-wedge boundary condition in the meridional direction. At the radial boundaries, the density, azimuthal velocity, pressure in the ghost cells are reset to their initial profiles, while the radial and meridional velocities and magnetic field are copied from the adjacent active cells. The dust radial velocity at the outer boundary is set to zero to prevent inflow of dust from outside the computational domain. Since our analysis is restricted to the inner region of the simulations ($R \lesssim 12$ and $|z| \lesssim 6$, see Section~\ref{sec:results}), the impact of the outer radial boundary condition on the disk dynamics is expected to be negligible.

\subsection{Numerical Diagnostics and List of Simulations}~\label{subsec:simulations}

We list several numerical diagnostics to characterize the turbulent activity, dust settling, and dust concentration in this study.

The radial transport of angular momentum is quantified using the dimensionless $\alpha$~\citep{ShakuraSunyaev1973}, the integral of Reynolds and Maxwell stresses normalized by pressure. The radially averaged $\alpha$ parameter is defined as
\begin{equation}
  \langle \alpha\rangle_R \equiv \frac{ \int_{R_\mathrm{min}}^{R_\mathrm{max}} \left[ \frac{ \int_{-z_\mathrm{tr}}^{z_\mathrm{tr}} (\mathcal{T}_\mathrm{Rey}+\mathcal{T}_\mathrm{Max} )\,dz }{ \int_{-z_\mathrm{tr}}^{z_\mathrm{tr}} P\,dz } \right] dR }{ \int_{R_\mathrm{min}}^{R_\mathrm{max}} dR }
\label{eq:alpha}
\end{equation}
where $ \mathcal{T}_\mathrm{Rey} \equiv\rho_\mathrm{g} \delta v_{\mathrm{g},R}\delta v_{\mathrm{g},\phi} $ is Reynolds stress, and $\mathcal{T}_\mathrm{Max} \equiv-B_R B_\phi/(4\pi)$ is the Maxwell stress associated with magnetic fields. The radial average is evaluated over the range $R_\mathrm{min}=3$ to $R_\mathrm{max}=10$. Azimuthal velocity fluctuations , $\delta v_\mathrm{g}\equiv v_\mathrm{g}-\langle v_{\mathrm{g}}\rangle_\phi$, cannot be obtained from our 2D axisymmetric simulations. We therefore calculate $\mathcal{T}_\mathrm{Rey}$ using deviations from the initial velocities, $\delta v_\mathrm{g}\equiv v_\mathrm{g}-v_{\mathrm{g},\mathrm{init}}$. The resulting stress includes both laminar and fluctuating motions. On the other hand, if we replace $v_\mathrm{g,init}$ by time-averaged velocities $\langle v_\mathrm{g}\rangle_t$, we find that the resulting $\alpha$ value changes by $\sim 10^{-4}$, suggesting that the laminar contribution is negligible.

Similarly, the vertical transport of angular momentum by the MHD disk wind can be quantified by the wind stress, $\mathcal{T}_\mathrm{wind}\equiv-B_z B_\phi/(4\pi)$. However, our study focuses on dust concentration driven by turbulence and wind-induced gas dynamics within the disk region. Since the role of wind stress in driving disk accretion has been extensively investigated in previous studies~\citep{BaiStone2017Hall,CuiBai2020}, we omit its analysis here to avoid redundancy.

We compute the radially averaged dust--gas scale height ratio:
\begin{equation}
  \left\langle \frac{H_\mathrm{d}}{H_\mathrm{g}}\right\rangle_R \equiv \frac{ \int_{R_\mathrm{min}}^{R_\mathrm{max}} \sqrt{ \frac{ \int_{-z_\mathrm{tr}}^{z_\mathrm{tr}} \left(\rho_\mathrm{d} z^2/H_\mathrm{g}^2\right) dz }{ \int_{-z_\mathrm{tr}}^{z_\mathrm{tr}} \rho_\mathrm{d} dz } } \, dR }{ \int_{R_\mathrm{min}}^{R_\mathrm{max}} dR }
\label{eq:HdopverHg}
\end{equation}
This ratio serves as a diagnostic of the strength of vertical turbulence and the efficiency of dust settling in the disk. Smaller values correspond to more efficient settling and a thinner dust layer, whereas larger values indicate stronger vertical turbulent stirring.

We adopt the Roche density criterion from~\cite{LiYoudin2021} as an approximate indicator for the onset of planetesimal formation:
\begin{equation}
  \rho_\mathrm{R} \equiv \frac{9}{4} \sqrt{2\pi} Q \rho_\mathrm{g} \simeq 180 \rho_\mathrm{g}
\label{eq:Roche}
\end{equation}
where $Q$ is the Toomre parameter~\citep{Toomre1964}, which is taken as $Q = 32$ in this study, yielding a Roche density of $\rho_\mathrm{R} \simeq 180\rho_\mathrm{g}$. Although self-gravity is not included in this study, the Roche density provides a useful reference for identifying regions where gravitational collapse would be expected. Based on Equation~\ref{eq:Roche}, the Roche density should in principle vary both spatially and temporally throughout the disk. However, since our simulations are scale free, we simplify the analysis by adopting a constant Roche density normalized to the local gas density.

We perform a suite of 2D axisymmetric simulations to explore how dust concentration in turbulent disks with MHD winds depends on magnetic, thermal, and dust-related parameters (see Table~\ref{table:runs}). Our fiducial run (``Fid'') adopts an initial midplane plasma beta of $\beta_{\rm pl,mid,init}=10^{4}$, an ambipolar Els\"asser number of $Am_{\rm disk}=0.3$ in the disk region, instantaneous thermal cooling ($\beta_{\rm th,disk}=10^{-6}$), a dust Stokes number of $St =0.1$, and an initial dust--gas surface density ratio of $Z_\mathrm{d} \equiv\Sigma_{\rm d}/\Sigma_{\rm g}=0.01$.

To assess the sensitivity to the initial poloidal magnetic field strength, we carried out two additional runs, ``B3'' and ``B5'', with midplane plasma beta values of $\beta_{\rm pl,mid,init}=10^{3}$ and $10^{5}$, respectively. The effect of AD was examined in run ``Am3'' by increasing the disk Els\"asser number to $Am_{\rm disk}=3$. Thermal effects were explored in run ``T1'' by enabling finite thermal cooling in the disk with $\beta_{\rm th,disk}=1$. Finally, the role of dust--gas coupling was investigated in run ``S2'', which adopts a smaller dust Stokes number of ${\rm St}=10^{-2}$ with $Z_\mathrm{d} = 0.02$. Unless otherwise stated, all other parameters are the same as in the fiducial run. All models presented in the main text include dust feedback (the second term in the right hand side of Equation~\ref{eq:gas_mom} and terms related to dust in Equation~\ref{eq:gas_erg}). For comparison, we also present a simulation without dust feedback in Appendix~\ref{app:nofb} to demonstrate the necessity of dust feedback in the dust concentration. We run each model for over $2000P_0$, where $P_0 = 2\pi/R_0$. The ``S2'' model is evolved longer, up to $2200P_0$.

\section{Results}~\label{sec:results}

\begin{figure*}[htp]
\centering
\includegraphics[scale=0.55]{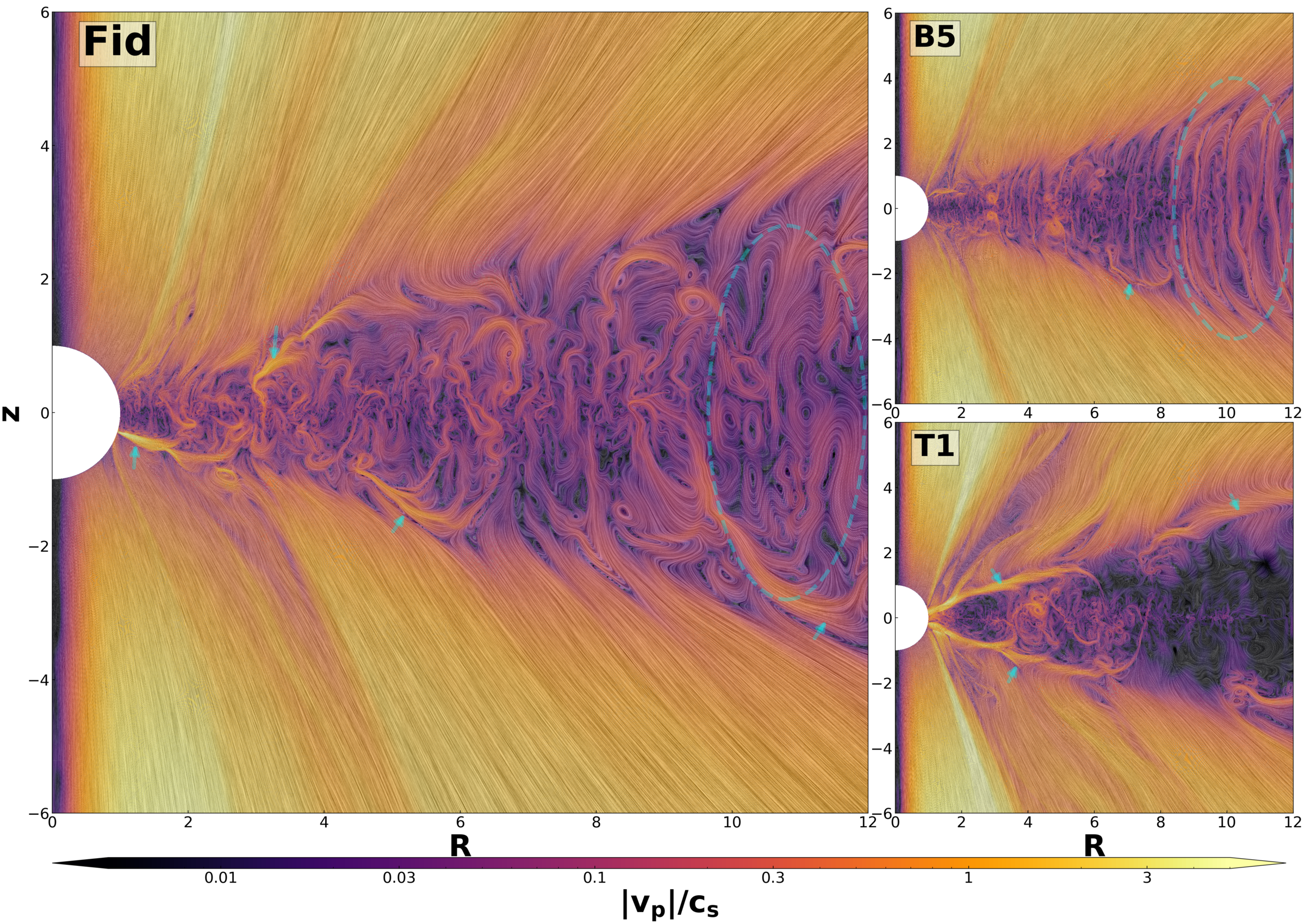}
\caption{Linear integral convolution (LIC) visualization of the poloidal (meridional) gas velocity field, normalized by the sound speed ($|\mathbf{v}_\mathrm{p}|/c_\mathrm{s}$), for the ``Fid'', ``B5'', and ``T1'' models at the final simulation snapshot ($t=2000P_0$). Model names are labeled in the top-left corner of each panel. The poloidal velocity structures in ``B3'', ``Am3'', and ``S2'' are qualitatively similar to those in ``Fid'' and are therefore omitted for clarity. The cyan arrows mark the locations of current sheets, while the cyan dashed ellipses highlight the VSI corrugation and higher-order body modes.}
\label{fig:lic_vel_p}
\end{figure*}

\begin{figure*}[htp]
\centering
\includegraphics[scale=0.30]{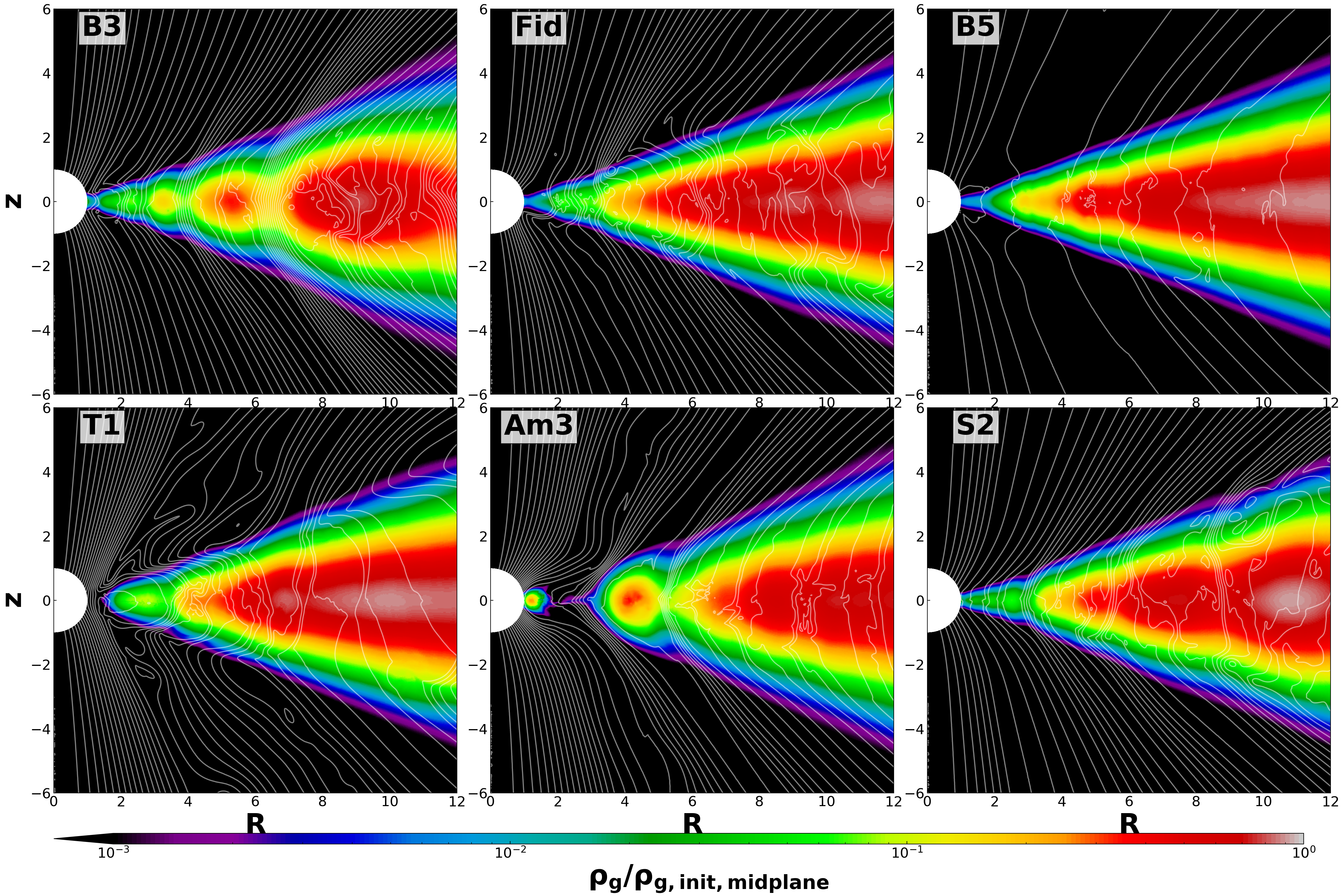}
\caption{The gas density normalized by the initial density profile in the midplane, $\rho_\mathrm{g}/\rho_\mathrm{g,init,midplane}$ at the final snapshots for different models (taken at $t=2000P_0$ for all models except ``S2'', which is shown at $t=2200P_0$). The contours indicate the poloidal magnetic flux.}
\label{fig:density_Bflux}
\end{figure*}

We present the results of this study here. Section~\ref{subsec:Property} describes the general gas properties of simulations. In Section~\ref{subsec:dust}, we investigate the resulting dust concentration and SI activity, while the evolution and concentration of magnetic flux are analyzed in Section~\ref{subsec:MFC}.

\begin{figure*}[htp]
\centering
\includegraphics[scale=0.60]{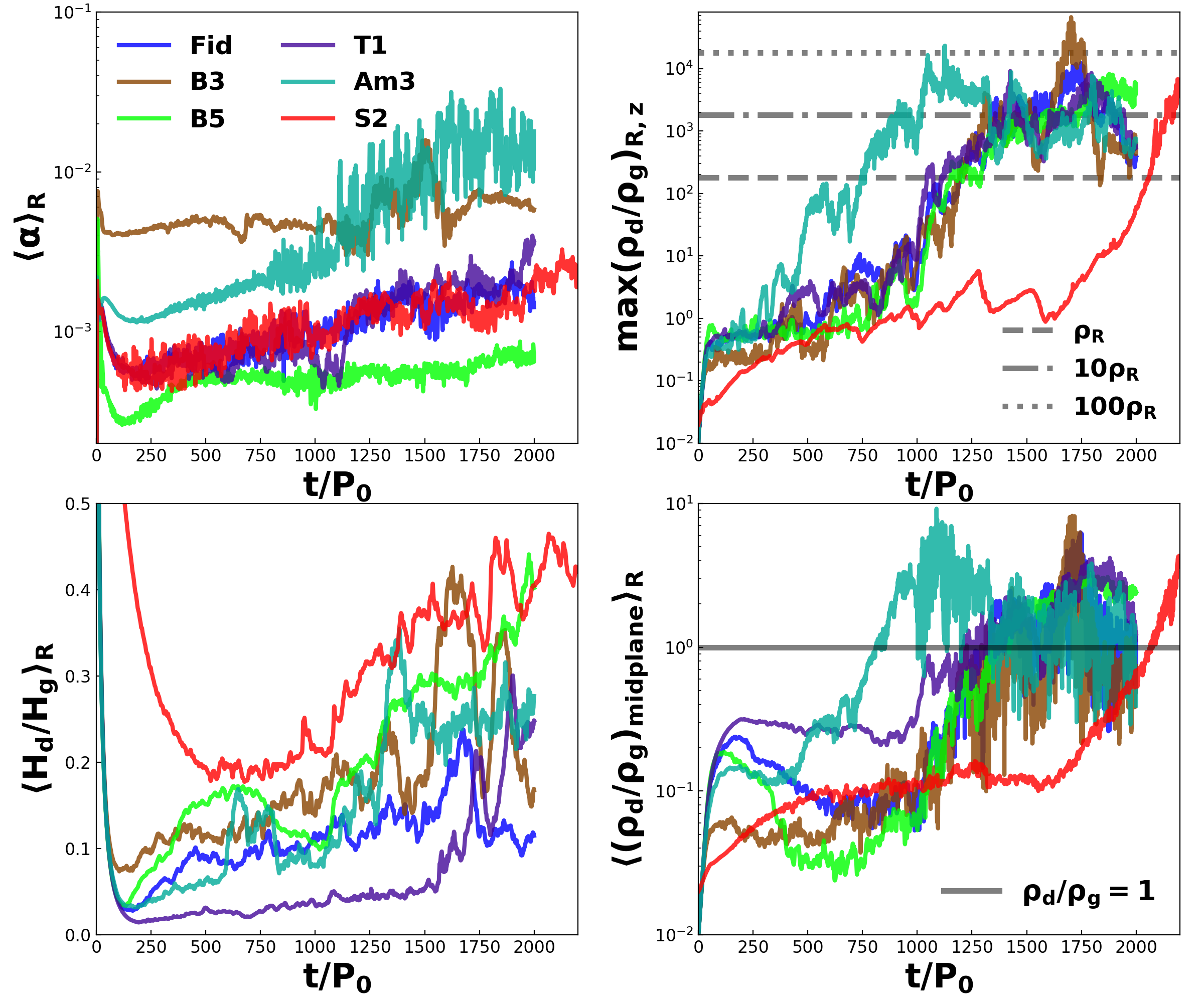}
\caption{From top to bottom and left to right, the panels show: the temporal evolution of the radially averaged $\langle \alpha\rangle_{R}$ (Equation~\ref{eq:alpha}); the temporal evolution of the maximum dust--gas density ratio, $\max(\rho_\mathrm{d}/\rho_\mathrm{g})_{R,z}$, where the maxima are computed within $R \in [3,10]$ and $z \in [-z_\mathrm{tr},\, z_\mathrm{tr}]$; the temporal evolution of the radially averaged dust--gas scale height ratio, $\langle H_\mathrm{d}/H_\mathrm{g} \rangle_R$ (Equation~\ref{eq:HdopverHg}); and the radially averaged dust--gas ratio in the midplane, $\langle (\rho_\mathrm{d}/\rho_\mathrm{g})_\mathrm{midplane} \rangle_R$ within $R\in[3,10]$. }
\label{fig:temporal_evolution}
\end{figure*}

\begin{figure}[htp]
\centering
\includegraphics[scale=0.40]{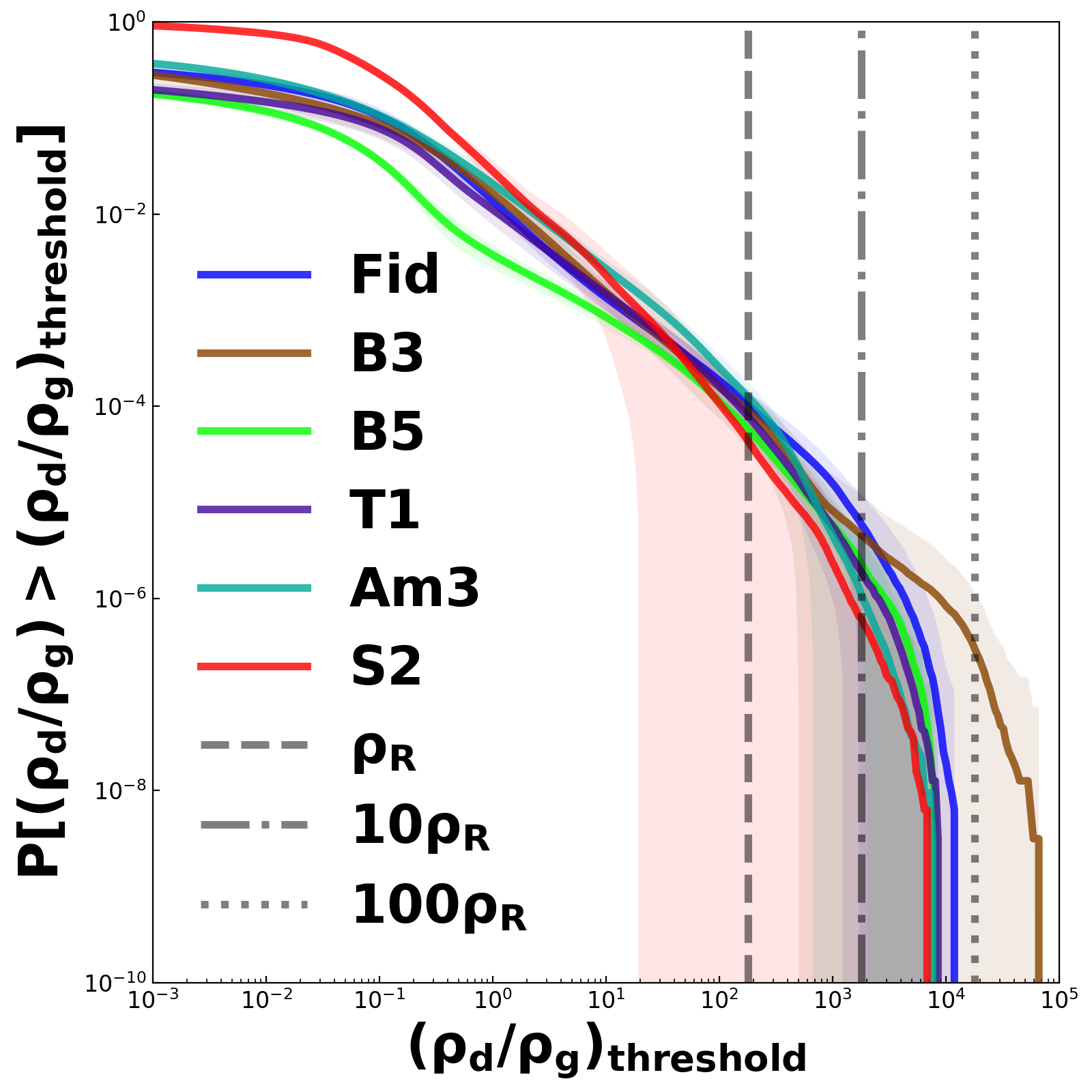}
\caption{The CDFs of the dust--gas density ratio, computed by counting the number of grid cells, $P[(\rho_\mathrm{d}/\rho_\mathrm{g}) > (\rho_\mathrm{d}/\rho_\mathrm{g})_\mathrm{threshold}]$. The CDFs are calculated using data from the last 500 innermost orbits ($1500P_0$--$2000P_0$ for all models except ``S2'', for which $1700P_0$--$2200P_0$ is used). The solid lines and shaped regions are the temporal averages and standard deviations. The calculation of the CDFs covers $R \in [3,10]$ and $z \in [-z_\mathrm{tr},\, z_\mathrm{tr}]$. For the CDF calculation across different refinement levels, cells on coarser levels are converted into an equivalent number of finest-level cells before performing the statistics.}
\label{fig:CDF}
\end{figure}

To facilitate the discussion in the following sections, we summarize the main diagnostics and their corresponding figures here. The overall gas dynamics and wind structure are illustrated in Figure~\ref{fig:lic_vel_p}, which shows the poloidal gas velocity normalized by the sound speed for representative models at final snapshots. Figure~\ref{fig:density_Bflux} presents the corresponding normalized gas density together with the poloidal magnetic flux structure for each model at final snapshots. The temporal evolution of key global diagnostics is summarized in Figure~\ref{fig:temporal_evolution}, including the $\langle\alpha\rangle_R$ parameter, the maximum dust--gas density ratio, the dust--gas scale height ratio, and the radially averaged dust--gas density ratio in the midplane. Figure~\ref{fig:CDF} shows the cumulative distribution functions (CDFs) of dust enrichment. Details of the CDFs calculation can be found in Section~3.3.3 of \citet{HuangBai2022}. The evolution of dust concentration is shown in Figures~\ref{fig:dust_ratio} and~\ref{fig:s-t-dust_ratio}. Figure~\ref{fig:dust_ratio} displays the dust--gas density ratio at the final snapshots for different runs, while Figure~\ref{fig:s-t-dust_ratio} shows the space--time ($R$--$t$) evolution of the maximum dust--gas density ratio. Finally, Figure~\ref{fig:s-t-B-rho} presents space--time diagrams of the vertical magnetic field and normalized gas density in the midplane, highlighting the coupling between magnetic processes and disk structure.

\subsection{Gas Properties}~\label{subsec:Property}

Variations in magnetic field strength mainly determine the relative importance of VSI and MHD activity in disk regions. Accordingly, we classify the six runs into three categories:
\begin{itemize}
\item \textbf{Intermediate-field cases}: ``Fid'', ``T1'', and ``S2''.
\item \textbf{Weak-field case}: ``B5''.
\item \textbf{Strong-field cases}: ``B3'' and ``Am3''.
\end{itemize}
Although ``Am3'' adopts the same initial poloidal field configuration as the intermediate-field cases, its weaker AD ($Am_\mathrm{disk}=3$) causes its magnetic dynamics to more closely resemble those of ``B3''.

\subsubsection{Radial Transport and Turbulence}~\label{subsubsec:transport}

Despite these differences among these models, the large-scale wind morphology remains qualitatively similar across the explored parameter space and is broadly consistent with previous non-ideal MHD wind simulations~\citep{BaiStone2017Hall}. Within the disk region (see Figure~\ref{fig:lic_vel_p}), gas motion remains subsonic, with a characteristic poloidal velocity of $|\mathbf{v}_p|\simeq 0.1 c_\mathrm{s} \sim 0.3\, c_\mathrm{s}$ (except ``T1'', which develops substantially smaller velocities), while magnetically driven outflows are launched above and below the transition layers ($|z|>z_\mathrm{tr}$) by the toroidal magnetic pressure gradient. The turbulent dynamics within the disk arise from the combined action of MHD activity, SI, and VSI (when present).

In the intermediate-field cases (``Fid'' and ``S2''), the VSI manifests as weak corrugation and higher-order body modes, most clearly visible around $R\simeq10$--$12$ (see the cyan dashed ellipse in the ``Fid'' panel of Figure~\ref{fig:lic_vel_p}), superimposed on stripe-like current sheets (cyan arrows in Figure~\ref{fig:lic_vel_p}) and magnetically driven meridional eddies. These current sheets are associated with reversals of the toroidal magnetic field (not shown; see, e.g.,~\cite{CuiBai2020}). Compared with previous locally isothermal simulations~\citep{Nelson2013}, the VSI-driven poloidal motions are relatively weak, as indicated by the small $|\mathbf{v}_p|/c_\mathrm{s}$ amplitudes in the ``Fid'' panel of Figure~\ref{fig:lic_vel_p}, while higher-order body modes become more prominent. This behavior is likely due to the combined effects of magnetic fields~\citep{CuiBai2020} and dust-induced buoyancy~\citep{LinYoudin2017,Lin2019,LehmannLin2022,HuangBai2025I}. These two models exhibit a moderate level of radial angular-momentum transport, with $\langle\alpha\rangle_R\sim10^{-3}$.

Thermal relaxation also plays a key role in regulating the VSI. In our intermediate-field case with a longer cooling time (``T1''), enhanced vertical buoyancy suppresses the VSI~\citep{Nelson2013,LinYoudin2015}, resulting in a more laminar disk structure. Consequently, the poloidal gas motion is the weakest among all models, with $|\mathbf{v}_p|\lesssim 0.03\,c_\mathrm{s}$ in the disk region (see the ``T1'' panel of Figure~\ref{fig:lic_vel_p}). Current sheets are nevertheless still present in ``T1'', indicating that the longer thermal relaxation time primarily suppresses the VSI while leaving the large-scale MHD activity qualitatively similar. Accordingly, the laminar component of the Maxwell stress continues to provide substantial angular momentum transport, such that $\langle\alpha\rangle_R$ remains of the same order as in ``Fid'', as shown in the upper-left panel of Figure~\ref{fig:temporal_evolution}.

In the weak-field model (``B5''), the disk dynamics more closely resemble the hydrodynamic limit. Specifically, the reduced magnetic field allows the VSI to develop most efficiently among all cases, producing pronounced corrugation modes (see the region highlighted by the cyan ellipse in the ``B5'' panel of Figure~\ref{fig:lic_vel_p}). Nevertheless, the radial angular-momentum transport remains the weakest among all models, with $\langle\alpha\rangle_R\simeq5\times10^{-4}$.

In contrast, the stronger-field models (``B3'' and ``Am3'') exhibit increasingly magnetically dominated gas dynamics. Compared with ``B5'', where radial angular momentum transport is strongly influenced by VSI-driven motions, the stronger magnetization in ``B3'' and magnetic coupling in ``Am3'' suppress the VSI and promote MFC (Figure~\ref{fig:density_Bflux} and Section~\ref{subsec:MFC}). In ``B3'', enhanced Maxwell stresses and laminar magnetic transport raise $\langle\alpha\rangle_R$ to $\sim5\times10^{-3}$, while the stronger magnetic contribution in ``Am3'' yields the largest $\langle\alpha\rangle_R$, reaching $\sim10^{-2}$ at late times.

\subsubsection{Density Depletion}~\label{subsubsec:depletion}

The MHD disk wind continuously extracts angular momentum with significant mass loss and drives secular disk evolution throughout the simulations. Generally, a stronger net poloidal field leads to faster wind-driven accretion and larger mass-loss rates~\cite[e.g.][]{BaiStone2016MTW}. As a result, a systematic reduction of gas density develops across the disk region. As shown in Figures~\ref{fig:density_Bflux} and~\ref{fig:s-t-B-rho}, the degree of gas depletion varies substantially in both space and time. Within the disk region of interest, the midplane gas density ratio, $\rho_\mathrm{g}/\rho_{\mathrm{g,init,midplane}}$, typically ranges from approximately $0.2$ to $0.8$ over $R\in[3,10]$, while substantially stronger local depletion occurs near the inner edge of this region ($R\simeq3$), where $\rho_\mathrm{g}/\rho_{\mathrm{g,init,midplane}}$ can fall below $0.1$. Consistently, the radially averaged gas surface density, $\langle\Sigma_\mathrm{g}\rangle_R$, measured over $R\in[3,10]$ and within $|z|<z_\mathrm{tr}$, decreases by approximately $30\%$--$48\%$ over $\sim2000$ innermost orbits (not shown).

The extent of depletion strongly depends on the magnetic field strength. The strong-field models (``B3'' and ``Am3'') exhibit the most gas loss, with $\langle \Sigma_\mathrm{g,final}\rangle_R \sim 30\% \,\langle \Sigma_\mathrm{g,init}\rangle_R$ and the formation of pronounced low-density regions associated with efficient angular momentum extraction and MHD disk winds. In contrast, the intermediate-field cases (``Fid'', ``S2'', and ``T1'') show moderate depletion with more spatially intermittent low-density structures, while the weak-field model (``B5'') retains the highest gas surface density among these models.

As there is little dust from the disk surface where wind is launched due to settling, this preferential gas removal naturally increases the local dust abundance, thereby providing favorable conditions for dust concentration (see next Section). In the meantime, we note that the wind mass-loss rates obtained in our simulations could be overestimated by up to a factor of several due to the simplified thermal treatment, particularly the lack of self-consistent radiative thermodynamics, as well as the inherent limitations of 2D axisymmetric simulations~\citep{WangBai2019}. Nevertheless, the scaling of the mass-loss rate with magnetic field strength remains robust across all models.

\subsection{Dust Concentration}~\label{subsec:dust}

\begin{figure*}[htp]
\centering
\includegraphics[scale=0.36]{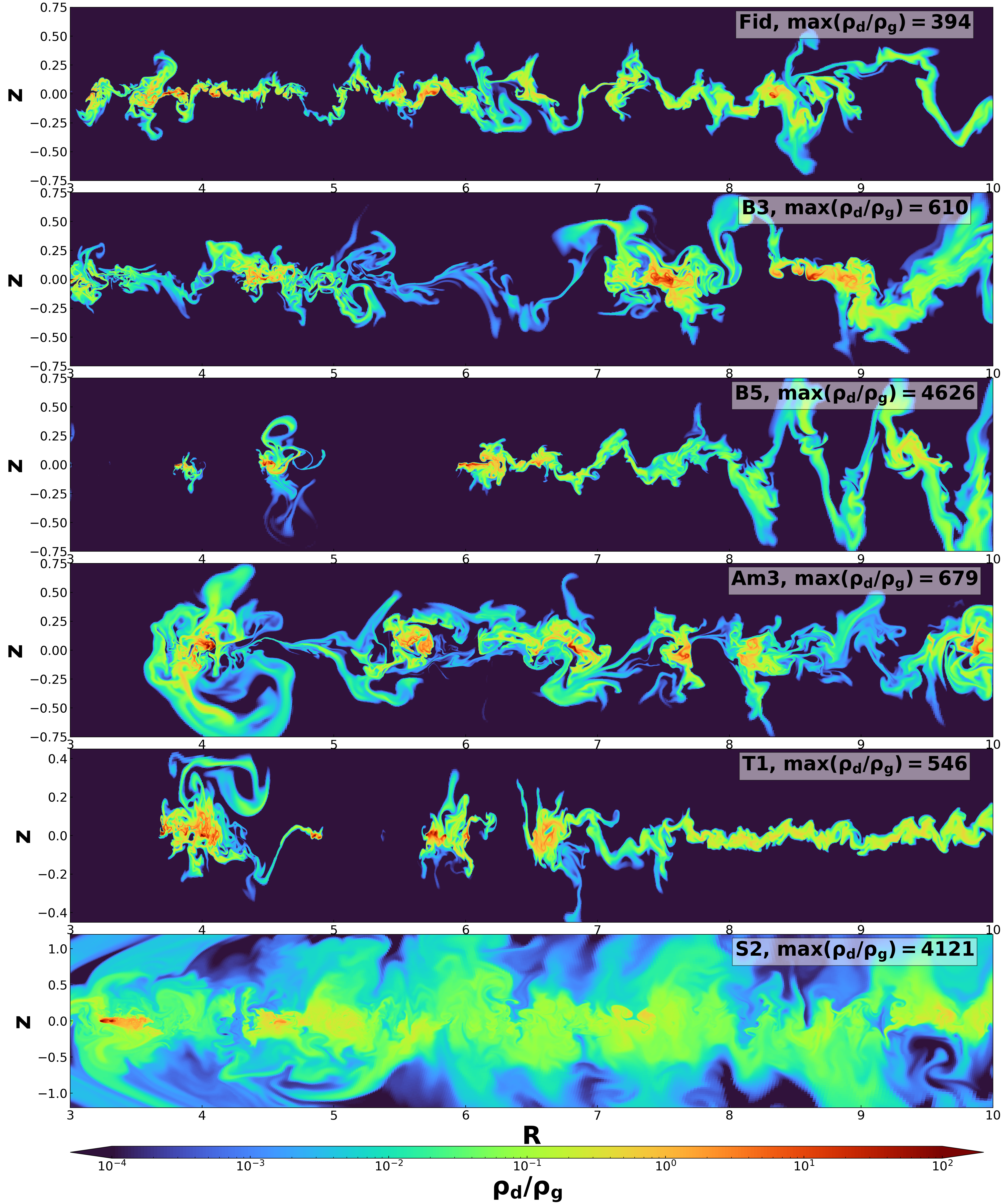}
\caption{dust--gas density ratios, $\rho_\mathrm{d}/\rho_\mathrm{g}$, at the final snapshots for different models. The model names and the corresponding maximum dust--gas density ratios within the radial range $R\in[3,10]$ are indicated in the top-right corner of each panel.}
\label{fig:dust_ratio}
\end{figure*}

\begin{figure*}[htp]
\centering
\includegraphics[scale=0.30]{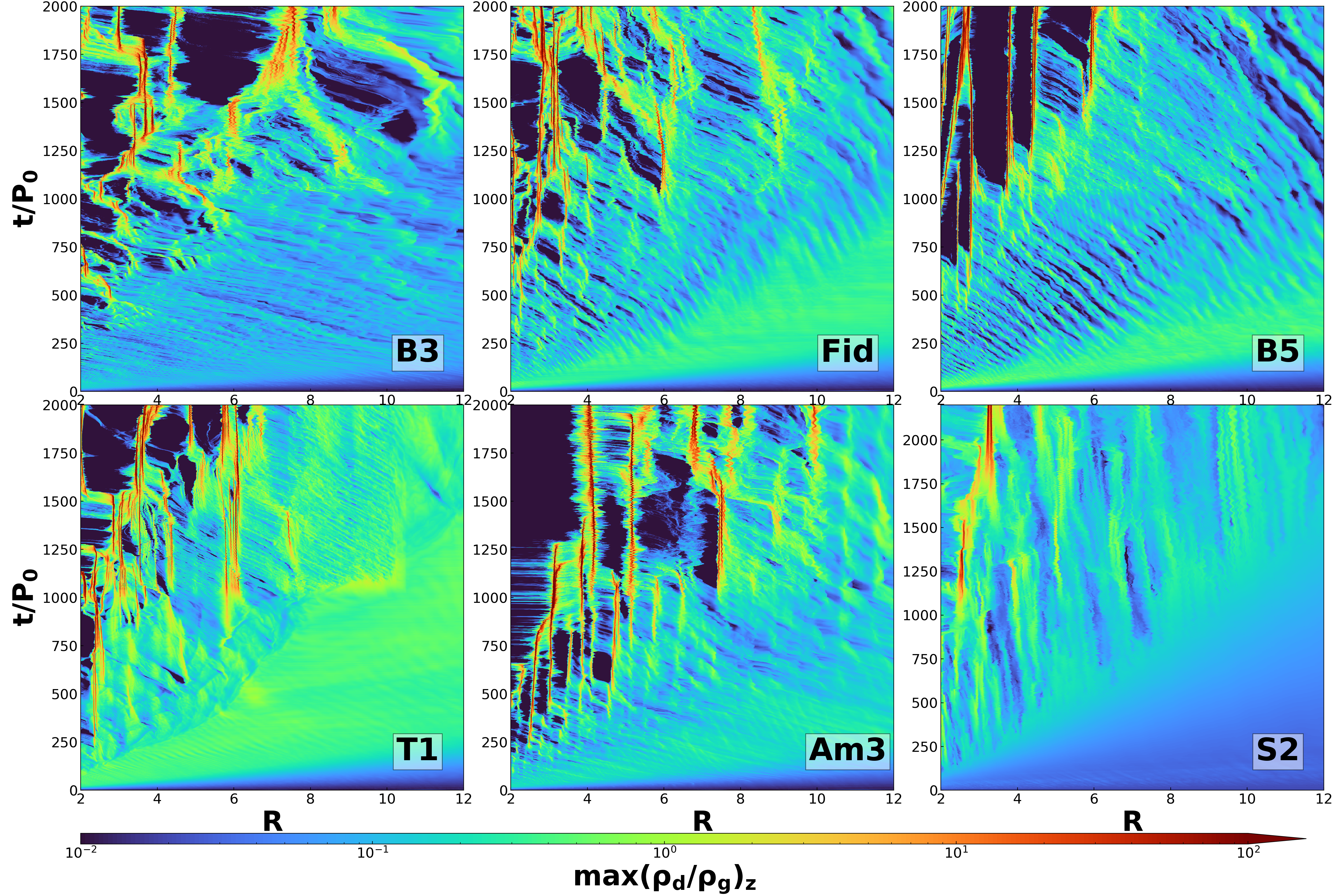}
\caption{Space-time $(R-t)$ plot for the maxima of dust--gas ratios $\max{(\rho_\mathrm{d}/\rho_\mathrm{g})_z}$ for different models.}
\label{fig:s-t-dust_ratio}
\end{figure*}

All models exhibit a common dust morphology characterized by strong dust concentration within our refinement region, with filamentary and intermittent clumpy structures driven by the combined effects of SI, VSI turbulence (when present), and MHD dynamics (Figure~\ref{fig:dust_ratio}). Compared with purely hydrodynamic cases~\citep{HuangBai2025I}, dust concentration is further enhanced by the gas depletion driven by MHD disk winds, which increases the local dust--gas ratio while leaving the dust distribution largely unaffected. As a result, the maximum dust concentration exceeds the Roche density $\rho_\mathrm{R}$ (Equation~\ref{eq:Roche}) in every model and reaches up to $\sim 10\rho_\mathrm{R} \simeq 1800\rho_\mathrm{g}$ during the simulations (top-right panel of Figure~\ref{fig:temporal_evolution}). Multiple long-lived dust clumps emerge in all runs (red spots in Figure~\ref{fig:dust_ratio}; see also the red streaks in Figure~\ref{fig:s-t-dust_ratio}). These clumps are not stationary at their formation locations but undergo radial drift under the background pressure gradient and can be disrupted by turbulent motions. Nevertheless, some clumps persist for more than $1000$ innermost orbits, indicating favorable conditions for sustained dust concentration and potential planetesimal formation.

It is customary to compare the conditions for dust clumping with the widely used SI clumping thresholds in dust abundance--Stokes number space identified by \citet{LiYoudin2021} and subsequent studies~\citep{LimSimon2024,LimSimon2026}. However, these criteria were derived from pure-SI simulations with a prescribed radial pressure gradient, commonly characterized by
\begin{equation}
\Pi \equiv -\frac{1}{2} \frac{H_{\mathrm{g,midplane}}}{R} \frac{d\ln P_{\mathrm{midplane}}}{d\ln R}
\label{eq:Pi}
\end{equation}
which is typically fixed at $\Pi=0.05$. The efficiency of SI-driven dust concentration is known to depend sensitively on the radial pressure gradient~\citep{Bai2010pressure,Baronett2024}. In this study, we adopt a substantially larger initial value, $\Pi=0.1625$, representative of the outer disk conditions considered here. Moreover, the local pressure gradient evolves dynamically in our simulations owing to MHD-driven disk evolution and zonal structures, further limiting a direct comparison with thresholds derived for a fixed $\Pi$. Therefore, the dust-abundance thresholds obtained in previous pure-SI studies cannot be directly applied to our simulations.

\citet{LimSimon2024} suggested that a mean midplane dust--gas density ratio exceeding unity provides a useful indicator of efficient SI-induced dust clumping in the saturated state. By comparing the top-right and bottom-right panels of Figure~\ref{fig:temporal_evolution}, we find that the onset of potential gravitational collapse (Equation~\ref{eq:Roche}) approximately coincides with $\langle (\rho_\mathrm{d}/\rho_\mathrm{g})_\mathrm{midplane}\rangle_R \gtrsim 1$ in our simulations (see the black dashed and solid lines). This consistency may suggest that the radially averaged midplane dust--gas density ratio can serve as a practical diagnostic for identifying the onset of dust clumping under more general conditions, such as VSI-driven turbulence or dust trapping in zonal flows induced by MFC (see Section~\ref{subsec:MFC}). Whether this threshold is universal or not remains to be established. Nevertheless, it provides a useful reference indicator that may be tested in future studies of dust concentration and planetesimal formation.

\subsubsection{Intermediate-field and weak-field cases: Fid and B5}~\label{subsubsec:FidandB5}

In the intermediate-field ``Fid'' model, dust concentration arises from the combined action of SI, VSI turbulence, and moderate MHD activity, leading to intermittent but sustained clump formation. In comparison, the weak-field ``B5'' model exhibits more wave-like dust structures due to the VSI~\citep{SchaferJohansen2020,HuangBai2025I}. In particular, stronger VSI corrugation modes continuously stir dust away from the midplane (Figure~\ref{fig:dust_ratio}), while SI-driven clumping still produces sustained dust concentration.

The global vertical dust distribution is characterized by the radially averaged dust--gas scale-height ratio, $\langle H_\mathrm{d}/H_\mathrm{g}\rangle_R$, which reflects the balance between dust settling and vertical stirring. Unlike $\langle\alpha\rangle_R$, $\langle H_\mathrm{d}/H_\mathrm{g}\rangle_R$ does not exhibit a simple dependence on magnetic field strength, which is primarily controlled by differences in vertical stirring. The weak VSI activity in ``Fid'' results in an evolution of $\langle H_\mathrm{d}/H_\mathrm{g}\rangle_R$ comparable to those in the stronger-field models ``B3'' and ``Am3'', whereas the stronger VSI in ``B5'' continuously lofts dust away from the midplane and produces a systematically thicker dust layer.

The CDFs further quantify the degree of local dust concentration (Figure~\ref{fig:CDF}). Despite the presence of VSI turbulence, efficient dust concentration persists in both ``Fid'' and ``B5''. In the ``Fid'' and ``B5'' models, respectively, approximately $10^{-4}$ and $5\times10^{-5}$ of the grid cells exceed the Roche density threshold $\rho_\mathrm{R}$, while approximately $6\times10^{-6}$ and $2\times10^{-6}$ exceed $10\rho_\mathrm{R}$. For comparison, in our previous 2D VSI+SI simulations~\citep{HuangBai2025I}, only $\sim10^{-5}$ of the grid cells exceeded $\rho_\mathrm{R}$. Although the numerical setups differ between the two studies and therefore prevent a direct quantitative comparison, the larger high-density fractions found here indicate that strong SI-assisted dust clumping remains effective in the present turbulent and magnetized disk environment.

\subsubsection{Strong-field cases: B3 and Am3}~\label{subsubsec:B3andAm3}

Dust concentration becomes stronger in the strong-field models ``B3'' and ``Am3''. MFC (Section~\ref{subsec:MFC}) generates pronounced radial pressure structures that develop into geostrophically balanced\footnote{The Coriolis force balances the radial pressure gradient.} zonal flows, thereby reducing radial drift and trapping dust. Although geostrophic balance itself is not specific to strongly magnetized disks, the stronger magnetic field produces steeper pressure bumps through MFC, thereby enhancing dust trapping. Near the pressure maxima, the radial pressure gradient and the associated radial dust--gas drift are reduced, which may locally weaken the conventional SI. In such regions, the azimuthal-drift streaming instability~\cite[ADSI;][]{LinHsu2022,HsuLin2022}, which does not require a radial pressure gradient, could in principle become more relevant. The ADSI requires the relative drift velocity between gas and dust in the background state to be primarily along the azimuthal direction. However, we find that the dust--gas drift to be dominated by the radial component over the bulk of the disk region.

The dust density of ``B3'' could exceed $10\rho_\mathrm{R}$ and even $100\rho_\mathrm{R}$. Previous studies have shown that dust feedback can further amplify dust accumulation within pressure bumps even when the conventional SI is weakened~\citep{XuBai2022Trapping}. Thus, the strong dust concentration in ``B3'' likely results from the combined effects of dust trapping by MFC-induced pressure bumps, dust feedback, and the continued activity of the conventional SI. The ``B3'' model exhibits the strongest dust concentration among all runs, reaching a peak value of $\rho_\mathrm{d}/\rho_\mathrm{g}\sim6\times10^{4}$ at $t\simeq1700\,P_0$. In the ``Am3'' model, dust concentration develops even earlier, exceeding $\rho_\mathrm{R}$ and $10\rho_\mathrm{R}$ at $t\simeq750\,P_0$ and $t\simeq900\,P_0$, respectively, prior to the activation of mesh refinement ($t=1000\,P_0$). This early enhancement is primarily driven by rapid gas depletion and MFC during the early evolutionary stages (Section~\ref{subsec:MFC}).

Despite their stronger magnetic fields, the dust-layer thicknesses in ``B3'' and ``Am3'' remain broadly comparable to that in ``Fid''. In ``B3'', $\langle H_\mathrm{d}/H_\mathrm{g}\rangle_R$ remains relatively elevated throughout the evolution and exhibits pronounced temporal fluctuations at late times. Since the VSI is strongly suppressed in the strong-field models, this vertical redistribution is instead associated with large-scale magnetically driven poloidal gas motions around the MFC-induced zonal structures and low-density gaps, which intermittently stir dust away from the midplane~\citep{RiolsLesur2020,HuLiZhu2022}. More specifically, in ``Am3'', the dust layer is initially comparable to or slightly thinner than that in ``B3'', but $\langle H_\mathrm{d}/H_\mathrm{g}\rangle_R$ increases relatively rapidly around $t\sim1250\,P_0$. It coincides with a pronounced depletion of dust over $R\simeq5$--$7$ (see Figure~\ref{fig:s-t-dust_ratio}), which modifies the radially averaged dust-layer thickness defined in Equation~\ref{eq:HdopverHg}. Similar episodic increases in $\langle H_\mathrm{d}/H_\mathrm{g}\rangle_R$ also occur in ``B3'' around $t\sim1500\,P_0$ and $1800\,P_0$, where comparable dust redistribution and depletion are present (Figure~\ref{fig:s-t-dust_ratio}). 

The CDF of ``B3'' exhibits the most extended high-density tail among all models, with approximately $3\times10^{-7}$ of the cells exceeding $100\rho_\mathrm{R}$, consistent with its sustained strong dust concentration and the presence of long-lived zonal flows. In contrast, the CDF of ``Am3'' shows a weaker high-density tail compared to ``B3'', but still deviates from the intermediate-field cases by exhibiting an enhanced high-density fraction relative to ``Fid'' and ``B5''.

\subsubsection{Intermediate-field cases with additional physical effects: T1 and S2}~\label{subsubsec:T1andS2}

In ``T1'', where VSI is suppressed, the disk becomes more laminar, yet SI remains active and still produces strong dust clumping, with the maximum dust density exceeding $10\rho_\mathrm{R}$. The reduced turbulent activity limits vertical mixing and leads to efficient dust settling. Consistently, ``T1'' exhibits the smallest $\langle H_\mathrm{d}/H_\mathrm{g}\rangle_R$ among all models, indicating a thin dust layer controlled by weak vertical stirring. The CDF of ``T1'' is similar to that of ``Fid'', reflecting comparable levels of SI-driven dust concentration.

In ``S2'', the smaller Stokes number ($\mathrm{St}=0.01$) results in tighter dust--gas coupling, which modifies the clumping behavior by reducing settling efficiency and maintaining more diffuse dust structures. Nevertheless, two prominent dust clumps develop at $R\simeq3.2$ after $t\simeq2000\,P_0$, when $\langle (\rho_\mathrm{d}/\rho_\mathrm{g})_\mathrm{midplane}\rangle_R \gtrsim 1$. The maximum dust density subsequently exceeds $10\rho_\mathrm{R}$ at $t\simeq2200\,P_0$, demonstrating that efficient SI-induced clumping can still occur despite the reduced settling efficiency. Consistent with its smaller Stokes number, ``S2'' exhibits the largest $\langle H_\mathrm{d}/H_\mathrm{g}\rangle_R$ among all models due to stronger dust--gas coupling. Its CDF is shifted toward lower dust--gas ratios ($\rho_\mathrm{d}/\rho_\mathrm{g}\lesssim10$), reflecting the tighter coupling. Nevertheless, about $4\times10^{-5}$ and $6\times10^{-7}$ of the cells exceed $\rho_\mathrm{R}$ and $10\rho_\mathrm{R}$, respectively, demonstrating that efficient dust concentration is still achieved even for tightly coupled grains.

Overall, dust concentration in these six models is regulated by the interplay between SI, turbulent activity, gas depletion driven by MHD disk winds (see Section~\ref{subsec:Property}), and MFC (see Section~\ref{subsec:MFC}), rather than a single dominant process.

\subsection{Magnetic Flux Concentration (MFC)}~\label{subsec:MFC}

Magnetic dynamics largely depends on the amount of poloidal magnetic flux threading the disk, as well as the evolution of its spatial distribution over time. In our simulations, all models exhibit a secular loss of poloidal magnetic flux due to AD~\citep{BaiStone2017Hall,Lesur2021}. Although Figure~\ref{fig:density_Bflux} shows only individual snapshots, traces of the secular flux loss can be seen in ``B3'' and ``T1'' at smaller cylindrical radii, where the equally spaced magnetic-flux contours become more widely separated. Quantitatively, the strongest relative flux reduction occurs in model ``B3'', whereas the flux relative loss is weakest in ``Am3'' and ``B5''. In addition to this global flux loss, a pronounced spatial redistribution of poloidal magnetic flux is observed in all models, leading to localized regions of MFC. In these regions, enhanced magnetic flux increases accretion and mass-loss rates, resulting in local gas depletion. MFC has been extensively reported in non-ideal MHD simulations of PPDs~\citep{BaiStone2014MFC,Suriano2018,RiolsLesur2020,HuLiZhu2022,CuiBai2021,HsuLi2025}. Enhanced MFC can subsequently lead to the formation of gas gaps, with associated zonal flows developing between the gaps, thereby providing favorable conditions for dust trapping and concentration.

\begin{figure*}[htp]
\centering
\includegraphics[scale=0.60]{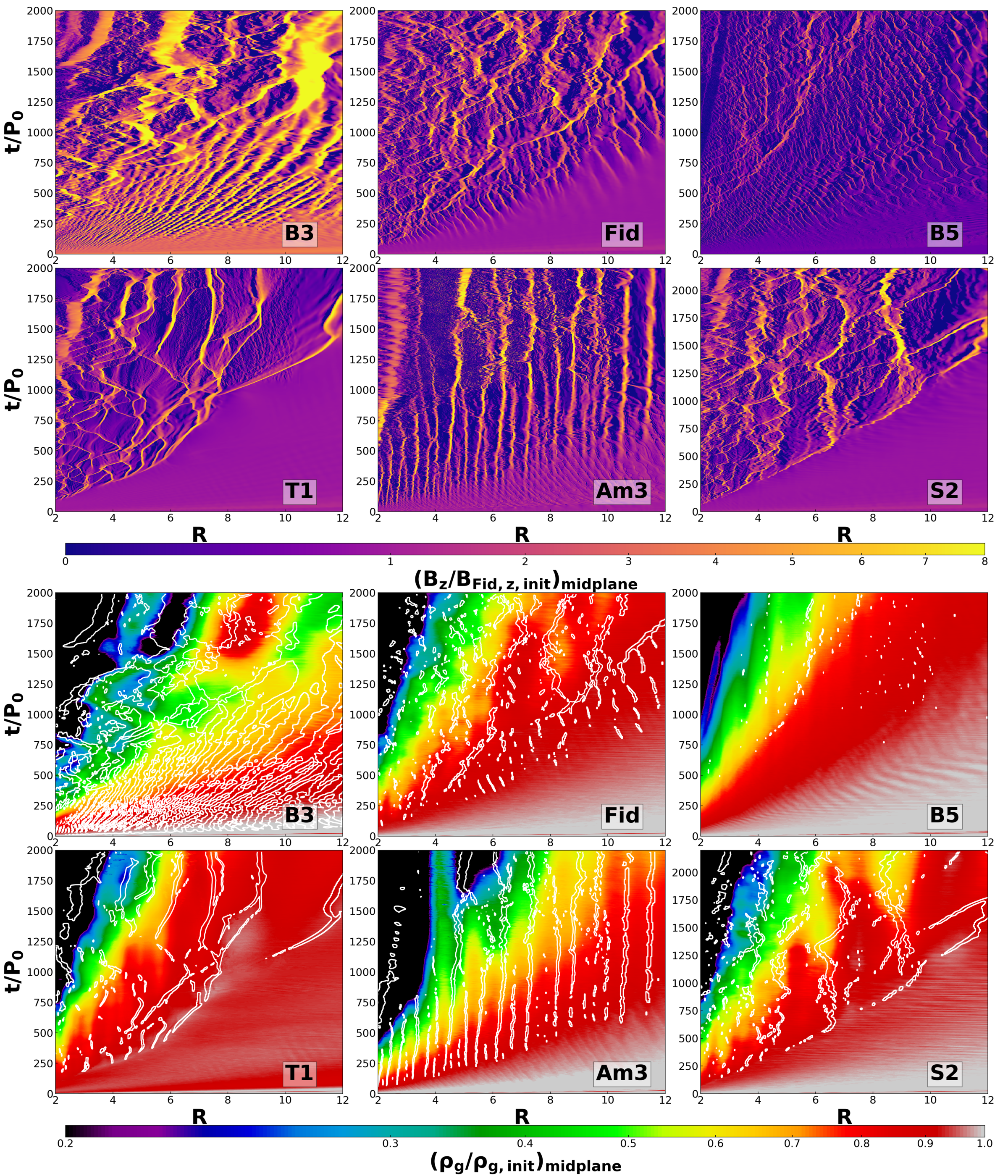}
\caption{The top six panels show space--time ($R$--$t$) diagrams of the vertical magnetic field in the midplane ($z=0,\theta=\pi/2$) for different models, normalized to the initial value in the fiducial run, $(B_z/B_{\mathrm{Fid},z,\mathrm{init}})_{\mathrm{midplane}}$. The bottom six panels show space--time ($R$--$t$) diagrams of the gas density normalized by the initial profile in the midplane,$(\rho_\mathrm{g}/\rho_{\mathrm{g},\mathrm{init}})_{\mathrm{midplane}}$. The white contours indicate the midplane vertical magnetic field normalized by its initial value in the fiducial run, $(B_z/B_{\mathrm{Fid},z,\mathrm{init}})_{\mathrm{midplane}}=3$. The contours are smoothed using a moving average with a window of $25P_0$ along the time axis.}
\label{fig:s-t-B-rho}
\end{figure*}

MFC exhibits a strong dependence on magnetic field strength. In the weak- and intermediate-field models (``B5'', ``Fid'', ``T1'', and ``S2''), MFC remains weak throughout the simulations and does not produce persistent gas gaps capable of efficiently trapping dust (Figures~\ref{fig:density_Bflux},~\ref{fig:s-t-dust_ratio}, and~\ref{fig:s-t-B-rho}). Instead, the disk dynamics are primarily governed by the combined effects of SI, VSI (except in ``T1''), and magnetically driven processes, including current sheets, large-scale poloidal gas motions (Section~\ref{subsubsec:transport}), and secular gas depletion driven by the MHD disk wind (Section~\ref{subsubsec:depletion}). In particular, ``B5'' represents the closest case to the hydrodynamic limit, where vigorous VSI activity persists and MFC is negligible. Although weak zonal flows are present in these models, they are not clearly associated with coherent MFC and are more likely generated by turbulence and its nonlinear interaction with magnetic fields. Consequently, dust concentration in the weak- and intermediate-field models is primarily associated with SI-driven or SI-assisted clumping in a background modified by VSI turbulence and magnetically driven disk evolution, with MFC contributing only marginally to the overall concentration process.

In contrast, in the strong-field model ``B3'', MFC becomes more efficient and produces well-identified radial structures. Clear gas gaps develop at $R\simeq3.5$, $6$, and $11$ at the final snapshot. These gaps show a clear spatial correlation with regions of enhanced magnetic flux (white lines in Figure~\ref{fig:density_Bflux} and white contours in Figure~\ref{fig:s-t-B-rho}). They are also associated with zonal flows located at $R\simeq2.5$, $5$, and $9$, which have radial widths of $5\sim 8H_\mathrm{g}$. These zonal flows act as long-lived pressure bumps and survive over 1000 innermost orbits. As a result, dust clumps are spatially correlated with these zonal flows, indicating that MFC-induced structures reduce radial drift and promote dust accumulation. MFC-induced zonal flows should therefore not be regarded as a pathway entirely independent of the SI. Pressure bumps can provide favorable sites for planetesimal formation by concentrating solids. ~\cite{Carrera2021} reported efficient SI-driven planetesimal formation for sufficiently large grains in pressure bumps at approximately solar metallicity. More relevant to the magnetized environment considered here, ~\cite{XuBai2022Trapping} found strong dust clumping in turbulent pressure traps using 3D non-ideal MHD simulations. In our simulations, MFC provides a self-consistent mechanism for generating such pressure structures, while the subsequent dust concentration can involve both trapping and SI-related processes, as discussed in Section~\ref{subsubsec:B3andAm3}.

A similar behavior is observed in ``Am3'', where MFC develops earlier ($t\simeq200P_0$) and produces prominent gaps at $R\simeq3$ and $R\simeq5$ by $t\simeq2000P_0$ (Figures~\ref{fig:density_Bflux} and~\ref{fig:s-t-B-rho}). MFC-induced zonal flows play a central role in dust trapping and in reducing radial dust drift. The number of dust clumps is comparable to the number of MFC-generated gaps, although the gap spacing is smaller than in ``B3'', reflecting the more compact radial structure of this model. SI-assisted clumping continues to occur in this slowly drifting dust population and produces additional small-scale structures. We note that $Am_\mathrm{disk}>1$ in this model, implying that MRI could significantly contribute to the disk dynamics~\citep{CuiBai2022}, although sustained MRI turbulence cannot be captured in our 2D axisymmetric simulations~\citep{LesurOgilvie2008MRI,LesurOgilvie2008dynamo}. Therefore, its contribution cannot be uniquely separated from other magnetically driven gas motions in the present simulations.

\section{Summary}~\label{sec:summary}

\subsection{Conclusions}~\label{subsec:conclusions}

In this work, we use the multifluid dust module of Athena++ to investigate dust concentration by resolving the small-scale SI in a more realistic global disk environment under AD and launching of MHD disk wind. We explore the effects of the magnetic field strength, thermal cooling, AD strength, and dust size in our models. Our simulations demonstrate that strong SI-assisted dust clumping remains robust across a broad range of global MHD disk conditions, while magnetic and hydrodynamic processes modify the background environment in which the SI operates.

\begin{itemize}

\item Across the explored range of magnetic field strengths, cooling conditions, and dust sizes, large-scale MHD winds persist and preferentially remove gas from the disk surface, thereby secularly increasing the dust abundance. The SI remains active in these wind-launching disks, indicating that AD and wind-driven disk evolution do not suppress strong SI-assisted dust concentration.

\item In weakly and intermediately magnetized disks ($\beta_\mathrm{pl,disk} \gtrsim 10^4$ with $Am_\mathrm{disk}\lesssim1$), the SI remains active across different thermodynamic conditions. It can coexist with MHD disk winds and, in rapidly cooling disks, with VSI-driven turbulence, leading to efficient dust clumping.

\item In disks with stronger magnetic fields ($\beta_\mathrm{pl,disk} \simeq 10^3$ or larger $Am_\mathrm{disk}$), magnetic effects become increasingly important: the VSI is suppressed, while MFC generates long-lived zonal flows that partially trap dust. With dust backreaction, the resulting dust concentrations can reach densities sufficient for planetesimal formation.

\end{itemize}

Overall, our simulations show that strong SI-assisted dust concentration can persist in PPDs with gas dynamics shaped by AD and large-scale MHD disk winds. Depending on the magnetic field strength and disk conditions, the background environment in which SI-related instabilities operate is modified by wind-driven gas depletion, SI--VSI interactions, and dust trapping in MFC-induced zonal flows. While individual contributions from these effects cannot be uniquely isolated in the present simulations, these results highlight the importance of global MHD models that self-consistently capture the coupled evolution of gas and dust.

\subsection{Caveats and Future Perspectives}~\label{subsec:caveats}

We have made several simplifying assumptions in this work, and future studies can extend beyond these limitations. First, we adopt 2D axisymmetric simulations. Such simulations tend to overestimate the turbulent energy in the vertical direction~\citep{JohansenYoudin2007,StollKley2014,HuangBai2025I}. Consequently, SI-driven dust clumping may be enhanced in 3D simulations~\citep{LimSimon2026}, where the vertical turbulent strength is generally weaker. Furthermore, 3D simulations allow non-axisymmetric modes and thus permit the development of additional instabilities, such as the Rossby wave instability~\citep{LiFinn2000,LiColgate2001,HsuLi2024} and the Kelvin--Helmholtz instability~\citep{Sekiya1998,Johansen2006,HuangBai2025II}. The non-linear saturation of VSI~\citep{MangerKlahr2018,HuangBai2025I,HuangBai2025II,Lesur2025} and MRI turbulence~\citep{CuiBai2021,CuiBai2022} may also differ in 3D simulations. Extending our simulations to 3D will therefore allow us to capture these effects and provide a more realistic description of disk dynamics.

Second, we adopt a simplified thermal treatment using a $\beta$-cooling prescription. The thermal structure of PPDs is regulated by stellar irradiation and radiative transfer processes \citep{ChiangGoldreich1997}. Since VSI turbulence is highly sensitive to disk thermodynamics, self-consistent radiation hydrodynamic simulations may yield different turbulence strengths and flow morphologies associated with the VSI \citep{StollKley2014,Flock2020,ZhangZhu2024}. Meanwhile, COS may also develop in intermediate cooling disks and influence dust concentration \citep{LehmannLin2025,LinLehmann2025,Klahr2026,KlahrBaehr2026}.

Third, we only include AD as the non-ideal MHD effect in the outer disk region. In the inner disk region, Ohmic resistivity and Hall effect may introduce more complex physical processes~\citep{Bai2014HallI,Bai2015HallII,Simon2015,Mori2025}.

Finally, we consider only a single dust species in our simulations. Including multiple dust species~\citep{Krapp2019,YangZhu2021SI,ZhuYang2021,Matthijsse2025} and dust coagulation~\citep{Tominaga2023,HoLiLi2024} may lead to different SI outcomes. In addition, our results could be further validated using particle--mesh methods~\citep{Bai2010particle,YangJohansen2016,MignoneFlock2019Dust} and different dust--gas codes~\citep{Baronett2026}. Exploring these effects will be an important direction for future work.

\begin{acknowledgments}
  We deeply grateful to the anonymous referee for the careful reading and constructive comments, which have substantially improved the quality and clarity of this manuscript. We thank for Mordecai-Mark Mac Low, Orkan Umurhan, Jim Stone, Min-Kai Lin, Zhaohuan Zhu and Rixin Li for helpful discussions. We also thank for Can Cui's problem generator as a template in this study. This work is supported by the National Science Foundation of China under grant No. 12503070, 12533011, 12325304 and 12233004. In this paper, we utilized the high-performance computing cluster ``Zimo'' from the Purple Mountain Observatory of the Chinese Academy of Sciences, which is equipped with GPU cards made in China.

\end{acknowledgments}

\software{Athena++~\citep{Stone2020,AthenaPP2024,HuangBai2022}}

\appendix

\section[]{A Test without Dust Feedback}\label{app:nofb}

We perform a test simulation without dust feedback to isolate its role in dust concentration. Since the gas does not feel the presence of dust in this case, we adopt a setup similar to the fiducial run but include two dust species with $St_1 = 0.1$ ($Z_\mathrm{d,1} = 0.01$) and $St_2 = 0.01$ ($Z_\mathrm{d,2} = 0.02$). We refer to this test as ``NoFB'' (No Feedback). The ``NoFB'' run is evolved for $2000 P_0$ to compare its results with those of ``Fid'' and ``S2'' in the main text.

\begin{figure*}[htp]
\centering
\includegraphics[scale=1.80]{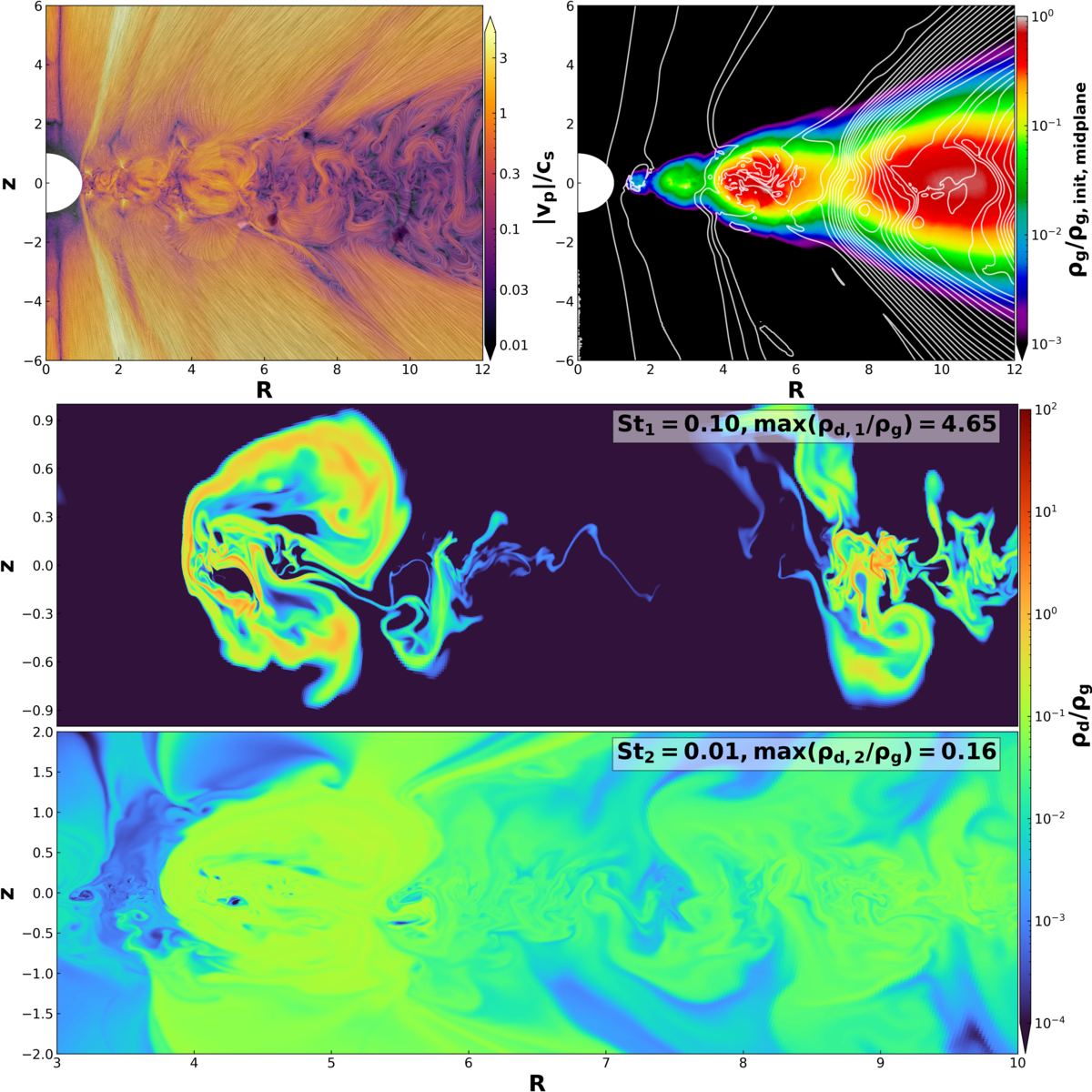}
\caption{The top-left panel shows the poloidal gas velocity normalized by the sound speed, similar to Figure~\ref{fig:lic_vel_p}. The top-right panel shows the normalized gas density together with the magnetic flux, similar to Figure~\ref{fig:density_Bflux}. The bottom panels present the dust--gas density ratios for two dust species ($St_1 = 0.1$ and $St_2 = 0.01$), similar to Figure~\ref{fig:dust_ratio}. The maximum dust--gas ratios for the two species are indicated in the top-right corners of the corresponding panels.}
\label{fig:nofb}
\end{figure*}

Figure~\ref{fig:nofb} shows the poloidal gas velocity, the normalized gas density, and the dust--gas density ratios at the final snapshot ($t = 2000\,P_0$). In the absence of dust feedback, VSI turbulence is noticeably stronger than in the ``Fid'' and ``S2'' models, whereas the wind motions above and below $\pm z_{\rm tr}$ remain comparable to those in the other models, since dust is largely absent in these wind regions.

Compared with ``Fid'' and ``S2'' (see Figure~\ref{fig:density_Bflux}), magnetic flux loss is stronger in ``NoFB'', particularly within $R\lesssim 6$. At the same time, pronounced MFC develops at $R\simeq 3.5$, $7$, and $12$, producing corresponding gas gaps at these locations. The stronger flux loss and MFC compared with those in ``Fid'' and ``S2'' suggests that magnetic flux transport is also influenced by dust mass loading and dust backreaction. Dust is subsequently trapped in the zonal flows associated with these gaps opened by MFC, resulting in two major dust concentration regions centered around $R\simeq 4.5$ and $R\simeq 9$.

Despite the presence of MFC-induced trapping, dust settling remains weaker than in ``Fid'' and ``S2'' for both dust species (see Figure~\ref{fig:dust_ratio}). By tracking the temporal evolution of dust concentration, we find that the maximum dust--gas density ratios only reach $\max(\rho_{\mathrm{d},1}/\rho_\mathrm{g})\simeq 40$ and $\max(\rho_{\mathrm{d},2}/\rho_\mathrm{g})\simeq 0.3$, remaining well below the Roche density required for planetesimal formation (Equation~\ref{eq:Roche}). This comparison demonstrates that dust feedback substantially enhances dust settling and concentration, and therefore plays an important role in facilitating planetesimal formation in PPDs.

\bibliographystyle{aasjournalv7}
\bibliography{references}{}

\end{document}